# The de Broglie' universal substratum, the Lochak' monopoles and the dark universe

Gilles Cohen-Tannoudji[a]
Laboratoire de recherche sur les sciences de la matière
(LARSIM, CEA, Université de Paris -Saclay)

ABSTRACT. We propose to interpret the dark universe (dark matter and dark energy) of the current cosmological standard model as emerging together with ordinary matter, from the quantum vacuum, which de Broglie called the "universal substratum". It will be shown that this interpretation may lead to a reasonable agreement between the current understanding of the quantum vacuum in quantum field theory and current evidences for the existence of dark energy and dark matter, and that dark matter can be interpreted as emerging from the QCD vacuum, as a Bose-Einstein gluon condensate, with an energy density relative to the baryonic energy density that agrees with observation.

## 1 Introduction

In a previous paper [1], a set of five foundational universal constants, the Planck's constant *h,* the vacuum celerity of light *c,* the Newton's gravitational constant *G,* the Boltzmann's constant $k_B$ , and the cosmological constant $\Lambda$, has been proposed to underly a general framework involving the interplay of three theories considering them, the quantum theory of fields (*h* and *c*), the quantum theory of information (*k* and c) and the general theory of relativity (*G, c,* and *Λ*). The outcome of [1] was that this framework considers the minimal number of foundational constants to allow, on a theoretical and metrological ground, the concordance of general relativistic and quantum physics. In the present paper, I intend to show that the cosmology, known as the cosmology of concordance, realizes, when interpreted in this framework the matching of the standard models of high energy physics (HEPSM) and of cosmology (CSM), and thus provides a way to extend *phenomenologically,* the concordance of general relativistic and quantum physics. After having summarized and made more precise the outcomes of [1], it will be shown that the cosmological standard model can be phenomenologically interpreted as the emergence of our quasi-classical observable universe, out of the quantum vacuum of the standard model of particle physics, without energy cost and in accordance with the requirements of general relativity and quantum physics. This approach is at odds with the current belief that the discovery of unexpected dark components of the content of the universe (dark matter and dark energy) would imply an irreducible contradistinction between the standard models of particle physics and general relativity. On the contrary, we think that it opens the possibility, if not of the complete unification of general relativity and quantum mechanics, at least, of their phenomenological quasi-classical concordance. In this prospect, I shall argue that the observed ratio of the cosmological parameter of the baryonic matter to the one of the dark matter can be made equal to $2N_f/11N_c$ where $N_c$ is the number of colors (equal to 3) and $N_f$ the effective number of light flavors of quarks (also equal to about 3) in the QCD sector of the standard model.

---

[a]Cet article fait suite à l'exposé présenté lors des Journés Louis de Broglie organisées le 4 juin 2019.
mailto:Gilles.Cohen-Tannoudji@polytechnique.org

I shall briefly conclude by an invitation to revisit two concepts that are essential for the de Broglie school of thought, namely, on the one hand, the quantum vacuum, that de Broglie called a "hidden subquantum medium" or a "universal substratum[b]", and the Lochak's monopoles [2], on the other hand.

---

[b] "The deepest level, *hypomicrophysical* or *sub-quantum* one could say, constituted by this "empty" immense reservoir of underlying energy of which we still know almost nothing. Words betray us to designate this deep level of reality: the word vacuum is not suitable at all because nothing would be fuller than this emptiness.

## 2 Quasi-classical cosmogony and its five foundational universal constants

What was called in [1] the *informational turn* of theoretical physics consists on the recognition of the foundational role played by information, and the inclusion in the panoply of foundational universal constants, of the Boltzmann's constant $k_\mathrm{B}$, that has the dimensional content of an entropy or an information. Moreover, when $k_\mathrm{B}$ is interpreted as a *quantum* of information and when the Planck's scale area $A_P = \mathrm{h}G / c^4$ is interpreted as a *quantum* of spacetime area, they underly, together with the *quantum* of action *ħ,* the putative general theoretical framework of a possible reconciliation of general relativistic and quantum physics, shown in figure 1 (taken, with slight modifications, from paper I).

What we call the domain of *quasi-classical phenomenological cosmogony* is the union of all the intersections of the three domains of the quantum theory of fields, of the quantum theory of spacetime and of the quantum theory of information, namely, their three double intersections and, in the center of the figure, their triple intersection, in which one can use *quasi-classical approximations*

---

The term "*universal substratum*" (or another of this kind) would be better. I will usually use the commonly used word vacuum, but you must imagine that it must be enclosed in quotation marks (the "vacuum")". Translated from *Lectures notes of the course by Louis de Broglie at La Sorbonne in 1957-58*

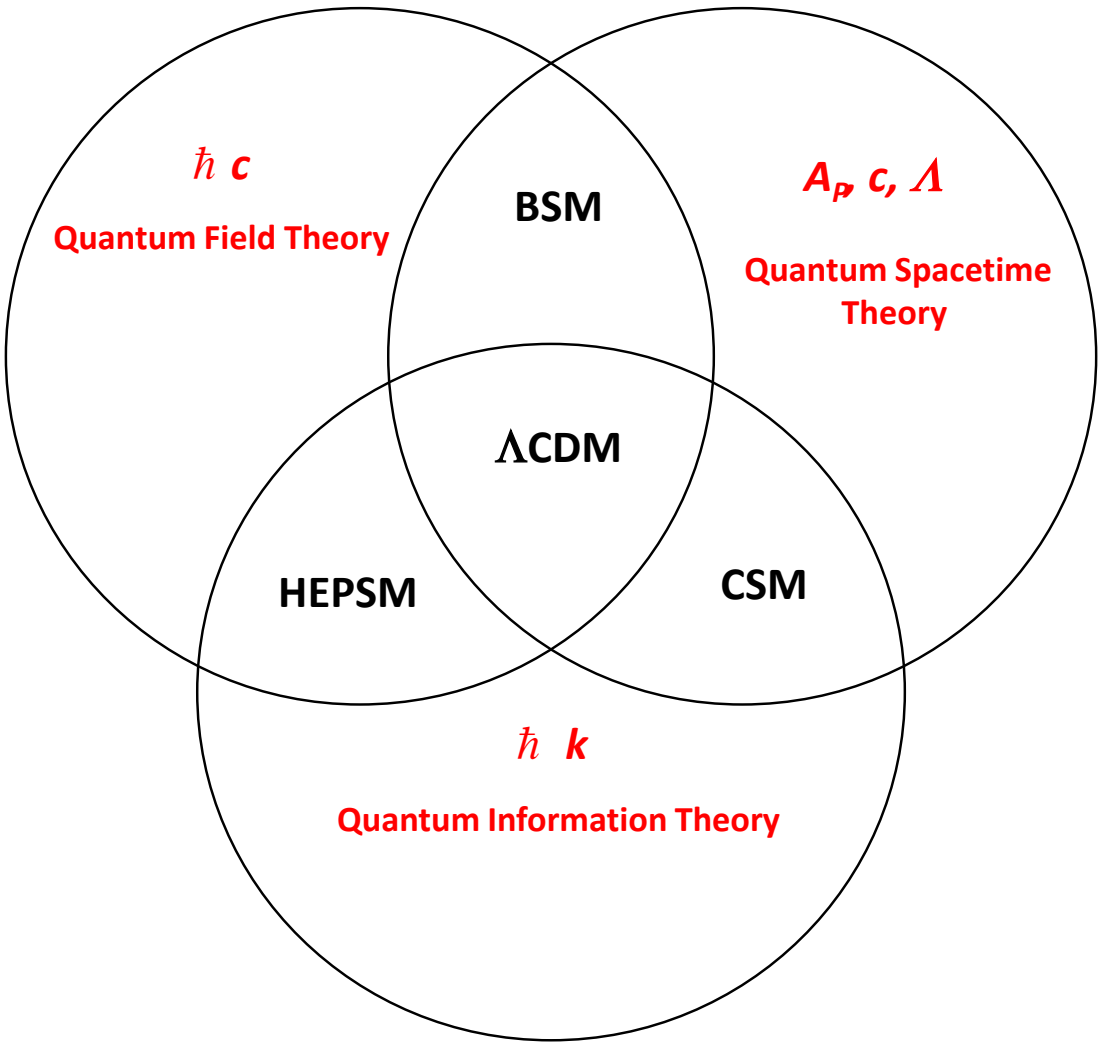

Figure 1

The fact that there is more than one elementary quantum allows one to develop some schemata, somehow generalizing the idea of complementarity [3], in which two quanta would go to zero with a fixed ratio depending on the third quantum and defining two complementary quasi-classical approximations. In the following sections, it will be shown that in each of the three double intersections in figure 1, such a schema can be at work allowing to solve the problems of the matching of the standard models of high energy physics and of cosmology:

1-Unruh's constant $\hbar / k_{\mathrm{B}}$ ratio and *Entanglement thermodynamics*;

2-Bekenstein' constant $k_{\mathrm{B}} \, / \, A_{\mathrm{P}}$ ratio and *Holography*;

3-Newton' constant $A_{\mathrm{P}} \, / \, \hbar$ ratio and *Gauge-gravity duality*.

I will show that the interplay of these three schemata allows raising ΛCDM, the cosmology of concordance, in the triple intersection, to the status of a *standard model of quasi-classical quantum cosmogony*.

## 3 Inflation, de Sitter (or anti de Sitter) spacetime, and emergence from the quantum vacuum

### *3.1 The three problems of hot big bang model (HBBM)*

The so-called HBBM, the first standard model of cosmology suffered from three problems linked to the primordial universe, the horizon problem, the flatness problem and the monopole problem.

The horizon problem stems from the fact that, in the HBBM, the primordial universe, is, as Gabriele Veneziano [4] says, "too large for its age": as a singularity, the big bang implies an expansion of space that is so rapid that different regions of the universe that can be observed today could not have been in causal contact in the primordial universe; now it turns out that all the CMB regions seem to be in thermal equilibrium.

The flatness problem is known as a "fine tuning problem": already in the sixties, the observations suggested that the density of the total content of the present-day universe is compatible with the so called critical density, namely the one corresponding to the vanishing of the spatial curvature index $k$, a time independent index; but maintaining such a vanishing throughout the history of the universe implies a huge fine tuning that prevents any reliable cosmological modelling.

The third problem of the HBBM is the absence of observation in the present-day universe of magnetic monopoles that are predicted to be produced in large numbers in the framework of non-abelian gauge theories unifying fundamental interactions.

### *3.2 The inflation scenario*

Historically it was soon realized that the main source of the difficulties of the HBBM was that the big bang is a *singularity* which prevents a causal description of the evolution of the primordial universe. The scenario of *inflation* was invented in order to avoid such a singularity. Inflation is a de Sitter phase of exponential expansion of the cosmic scale factor occurring in the primordial universe and increasing all the scales in such a way that all the three difficulties of the HBBM are solved provided that the duration of inflation is enough. The singularity of the big bang is avoided if the origin of the inflation phase is *a fluctuation of the quantum vacuum*, which means that the "Universe is spontaneously created from *nothing*" [5] In the rest of the present section, it will be shown that, in the five constants' theoretical framework proposed in [1], the new standard model of cosmology can be thought as realizing this program.

### *3.3 The quantum vacuum as a Minkowskian, cosmological background*

The idea in order to perform the matching of the standard models of particle physics and cosmology, is to interpret the quantum vacuum of the standard model of particle physics, namely the state of the Fock space for which all the occupation numbers of the involved quantum fields vanish, as the cosmological background underlying the standard model of cosmology, namely the spacetime metric of the dark components of the observable universe. The cosmological principle implies that in homogeneous and isotropic cosmologies such as ΛCDM that are *conformally flat* [6], the metric of the cosmological background is Minkowskian up to a multiplicative factor involving a scalar field $\phi$ called the *dilaton,* related to its determinant $|g|$

$$ds^2 = f(x)\left(ds\right)^2_{\text{Minkowski}} ; |g| = f(x)^4 |g|_{\text{Minkowski}} \qquad f(x) \propto \phi^2 \tag{1}$$

## 4 The assets of ΛCDM, the new standard model of cosmology

### *4.1 The concept of world-matter*

The starting point of modern cosmology considering the expansion of the universe and the possible existence of a cosmological constant is the Einstein's equation, which, following the definitions and notations of reference [7] and the proposal of Gliner [8] and Zeldovich [9] to take the cosmological term to the right-hand side, reads:

$$\mathcal{R}_{\mu\nu} - \frac{1}{2} g_{\mu\nu} \mathcal{R} = 8\pi \mathrm{G}_N T_{\mu\nu} + \Lambda g_{\mu\nu} . \tag{2}$$

The Robertson-Walker metric allows to describe a homogenous and isotropic universe by means of a metric, solution to the Einstein's equation, in terms of two cosmological parameters: the spatial curvature index $k$, an integer equal to -1, 0 or +1 and the overall dimensional expansion (or contraction) radius of the universe $R(t)$, depending only on time:

$$ds^2 = dt^2 - R^2(t)\left[\frac{dr^2}{1-kr^2} + r^2\left(d\theta^2 + \sin^2\theta d\phi^2\right)\right] \tag{3}$$

One often uses a dimensionless scale factor $a(t) = R(t)/R_0$ where $R_0 \equiv R(t_0)$ is the radius at present-day.

In order to derive the Friedman-Lemaitre equations of motion, one assumes that the matter content of the universe is a perfect fluid for which the energy momentum tensor is expressed in terms of the isotropic pressure $P$, the energy density $\rho$, the space time metric $g_{\mu\nu}$ and of the velocity vector $u = (1,0,0,0)$ for the isotropic fluid in co-moving coordinates

$$T_{\mu\nu} = -Pg_{\mu\nu} + (P+\rho)u_\mu u_\nu \ . \tag{4}$$

The cosmological term, taken to the right-hand side of Eq. (2) plays the role of what de Sitter called the energy density of a *world-matter*, namely, a contribution to the stress energy-tensor in the right-hand side of the equation proportional to the metric field $g_{\mu\nu}$. If, as in Eq. (4), one associates a perfect fluid to such a cosmological term, its isotropic pressure $P_\Lambda$ and energy density $\rho_\Lambda$ sum to zero and the world-matter energy tensor of the vacuum reduces to the pressure term, the sign of which is opposite to the one of $\Lambda$. A world-matter with a positive CC can be called a *de Sitter world-matter* (positive energy density, negative pressure) and a world-matter with a negative CC (negative energy density, positive pressure) can be called an *anti-de Sitter world-matter.*

### *4.2 The three stages of cosmic evolution according to ΛCDM*

In terms of the pressure and the density of the perfect fluid describing matter, the Friedman equations, with a non-vanishing CC, reads

$$H^2 \equiv \left(\frac{\dot{R}}{R}\right)^2 = \frac{8\pi G_N \rho}{3} - \frac{k}{R^2} + \frac{\Lambda}{3} \tag{5}$$

and

$$\frac{\ddot{R}}{R} = \frac{\Lambda}{3} - \frac{4\pi G_N}{3}\left(\rho + 3P\right). \tag{6}$$

Energy conservation (via $T^{\mu\nu}_{;\mu} = 0$) leads to a third equation:

$$\dot{\rho} = -3H\left(\rho + P\right). \tag{7}$$

The phases of the cosmic evolution are well represented on the thick line in figure 2 in which the Hubble radius $L = H^{-1}$, the inverse of the time dependent Hubble parameter, is plotted versus the scale parameter *a(t)* (set to 1 today) in logarithmic scale. On this figure, one can distinguish three stages: the *primordial inflation, the expansion and the late inflation*

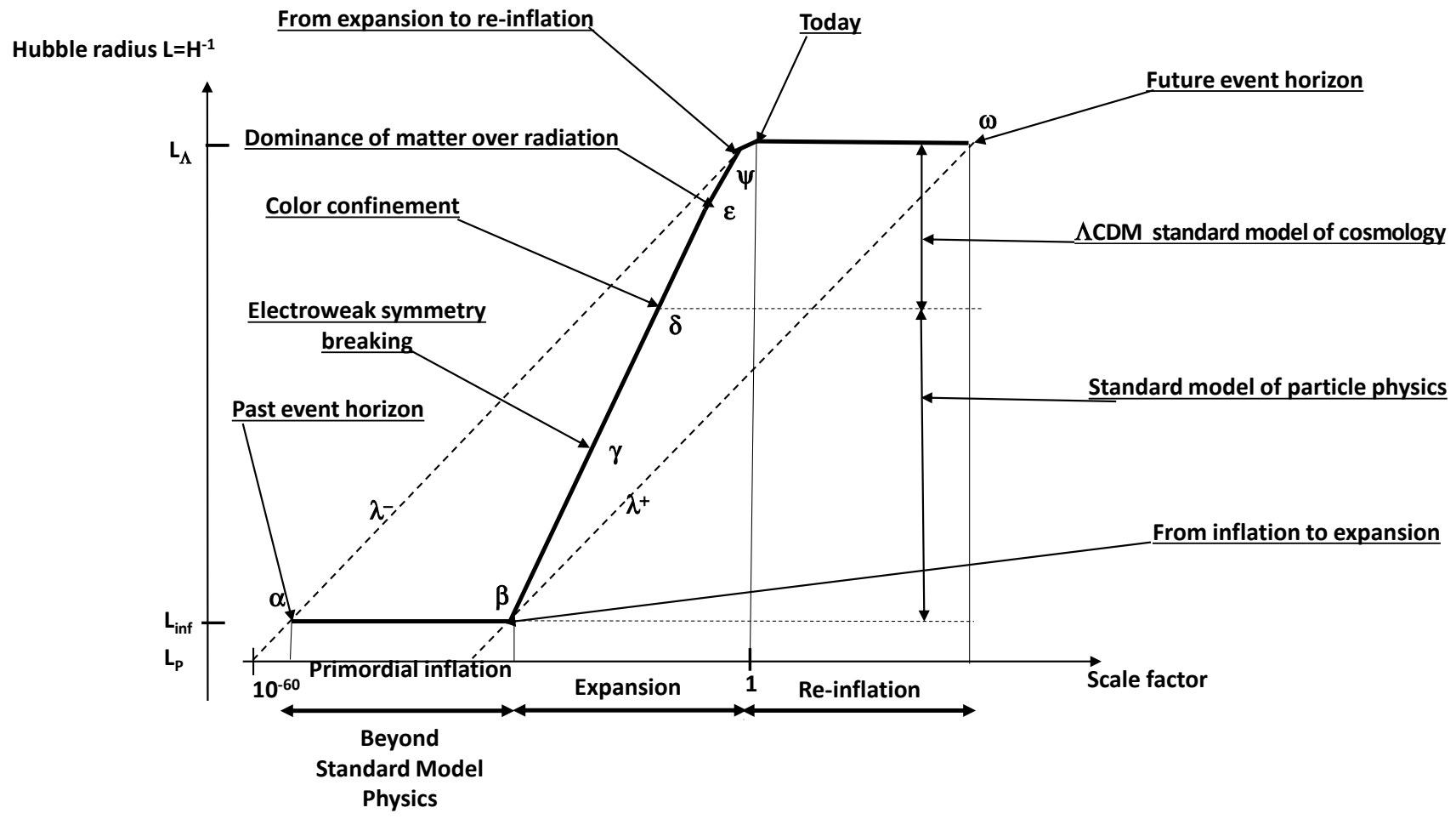

Figure 2 As explained in [1], the ΛCDM cosmology is represented in this graphic in which the Hubble radius $L = H^{-1}$ is plotted versus the scale factor $a(t)$ in logarithmic scale. The cosmic evolution is schematized on the thick line, on which the cosmic time grows linearly in the inflation phases (horizontal parts from point α to β and from ψ to ω) and exponentially in the expansion phase (from β to ψ). All quantum fluctuations exit from the Hubble horizon in the primordial inflation phase, enter it in the expansion phase and re-exit it in the late inflation phase. No information carrying quantum fluctuation with a wavelength smaller than $\lambda^-$(on the left of the *past event horizo*n α), or with a wavelength larger than $\lambda^+$ (on the right of the *future event horizon* ω) enters the Hubble horizon. From this limitation, one infers that the total content of information in the observable universe, which is proportional to the inverse of the cosmological constant is **finite**.

### *4.2.1 The primordial inflation*

The first stage of the evolution, called *primordial inflation* is a first de Sitter (or rather an anti-de Sitter) stage, occurring at an energy of about $10^{16}$ GeV, during which the horizon radius is constant (about $10^3$ Planck lengths) whereas the scale factor grows exponentially from point α to point β in figure 2 by about thirty orders of magnitude. During such a primordial inflation phase, the "young universe" [4] would not be anymore "too large for its age", the whole content of the universe would be in causal contact, which would solve the horizon problem. Inflation would also solve the flatness problem: it would flatten space in such a way that one can assume that at its end the spatial curvature is already compatible with zero, as confirmed by the agreement of the vanishing of the spatial curvature index with the present-day observations $\left(\Omega_k \equiv -k / R_0^2 H_0^2 = 0.0008^{+0.0040}_{-0.0039}\right)$ [10].

The fact that the scale at which this primordial inflation is compatible with the one at which Grand Unified Theories (GUT) are expected to be at work suggests that it could also allow solving the third problem of the HBBM, the one of the absence of magnetic monopoles that would be predicted to be produced in large numbers in the GUT framework: inflation would expel the GUT magnetic monopoles beyond the horizon.

### *4.2.2 The expansion*

At point β, primordial inflation is supposed to end and occurs, when the radius of expansion *R*(t) reaches about $10^{30}$ Planck's lengths corresponding to the energy scale of CC or to the Compton wave length of a particle with a mass of about 2.5 meV, what historically was called the *reheating* transition of the universe, and plays the role of a physical (i.e. non-singular) big bang.

Starting at point β, the second stage is an *expansion stage* during which, as in the hot big bang model, the content of the universe obeys the standard Friedman-Lemaître cosmological equations of evolution, with a time dependence of the cosmic scale $a(t)$ determined by the equation of state parameter $w = P / \rho$ of the component that dominates the evolution at a given epoch, namely;

- An epoch of dominance by radiation $(w = 1/3)$ from point β to point ε in figure 2 in which $L \propto a^2$ followed by,
- An epoch of dominance by pressure-less matter $(w = 0)$ from point ε to point ψ that corresponds to the transition from expansion to re-inflation in which $L \propto a^{1.5}$.

### *4.2.3 The late inflation*

In the third stage, extending from point ψ to point ω in figure 2, the universe will be dominated by the cosmological constant Λ. This stage, like the first one (from point α to point β) is a de Sitter stage, called re-inflation or *late inflation* with a scale factor growing exponentially with the cosmic time, and a Hubble radius slowly increasing asymptotically to $\sqrt{3/\Lambda}$ ) characterized by an equation of state $w=-1$ compatible with the present day observation, $\left(w=-1.019^{+0.075}_{-0.080}\right)$ [11]

### *4.3 The flatness sum rule*

It is useful to define a density, called the *critical density*

$$\rho_c \equiv \frac{3H^2}{8\pi G_N} , \tag{8}$$

which would be a solution to the Friedman's equation (5) if the curvature index *k* and CC were zero. Note that this critical energy density is just the negative isotropic pressure exerted on the boundary of the observable universe, the *Hubble horizon* (a 3-sphere with a radius equal to the Hubble radius).

With respect to this critical density one defines for each component, including the one of CC, its relative contribution to the critical density, called its dimensionless cosmological parameter, and rewrite the present day Friedman's equation (5) as

$$\begin{gathered} \Omega_{\text{tot}} = \rho / \rho_c \\ k / R^2 = H^2\left(\Omega_{\text{tot}} - 1\right) \\ k / R_0^2 = H_0^2\left(\Omega_{\text{M}} + \Omega_{\text{R}} + \Omega_{\Lambda} - 1\right) \end{gathered} \tag{9}$$

where the subscript M stands for pressure-less matter, the subscript R stands for radiation (or relativistic particles) and $\Omega_\Lambda = \Lambda / 3H^2$. Since the curvature index *k* does not depend on time, its vanishing at present day implies its vanishing at all epochs, so in terms of time-dependent densities, the Friedman's equation (5) takes the form of the *flatness sum rule* which, in terms of densities reads

$$\rho_{\mathrm{M}} + \rho_{\mathrm{R}} + \rho_{\Lambda} = \rho_c \qquad (10)$$

It appears that this flatness sum rule cannot be satisfied if, for the pressure-less matter energy density, one considers only the visible (baryonic) matter $\rho_{\mathrm{b}}$ unless one adds to it an unknown component of matter, called *dark matter*, $\rho_{\mathrm{DM}}$ – the adjective 'cold' meaning that it would contribute as a pressure-less matter ( $w = 0$ ), and one associates with the cosmological term $\rho_{\Lambda}$ a *dark energy* ( $w = -1$ ) component. In terms of densities, the Friedman's equation thus becomes

$$\begin{gathered} \rho_{\mathrm{b}} + \rho_{\mathrm{R}} + \rho_{\mathrm{DM}} + \rho_{\mathrm{DE}} - \rho_{\mathrm{c}} = 0 \\ \text{with } \rho_{\mathrm{DE}} = \rho_{\mathrm{V}} = \Lambda / 8\pi G_N \end{gathered} \qquad (11)$$

### *4.3.1 The lensing potential and a possible interpretation of dark matter as a world-matter*

The analysis in the Planck's experiment of the gravitational lensing affecting the travel of light from the emission of the cosmic microwave background to its measurement by the Planck satellite, favors an interpretation of dark matter as a component of a world-matter. The gravitational lensing blurs the map of the CMB, but thanks to the precise measurements of temperature and polarization correlations, it has been possible to unfold the effects of this blurring and thus, on the one hand, to recover the map of the original, un-blurred CMB, and, on the other hand, to draw the map of the lensing potential that has affected the light in its travel, which is nothing but the gravitational potential induced by the material content of the universe, dominated by baryonic matter. As a gravitational potential, its negative average energy can be interpreted as the one of an *anti-de Sitter component of the world-matter* (negative energy density, positive isotropic pressure), which can possibly be associated with dark matter, the other component (with negative pressure) being associated with dark energy.

*4.3.2 The problem of time and its solution according to the thermal time hypothesis*

But if one wants this interpretation to be at work not only at the present time but also at all epochs, one faces an apparent paradox: the dark energy density would be *constant*, whereas the dark matter would not be constant. But this is only the well-known "problem of time": how is time defined in cosmology?

To describe the cosmic evolution following the transition at point β in terms of a thermodynamical partition function (involving the quantum of entropy, $k_B$ rather than the quantum of action *h*), one must perform an analytic continuation from the Minskowskian spacetime of the path integral of quantum field theory to a Euclidean spacetime. The time of this Euclidean spacetime (an imaginary space dimension) is the *thermal time,* of which A. Connes and C. Rovelli [12] have shown that it can provide a way out of the cosmological problem of time. This thermal time $\tau$ is related to the geometrical time *t* through the Unruh's constant $U = \frac{\hbar}{k_B c}$ leading to a thermal bath with a temperature $T = U\frac{a}{2\pi}$ which is the temperature measured by a thermometer moving with an acceleration *a*: $T = a / 2\pi k_B$ : $\tau = \frac{2\pi}{a} t$ and

> "based on the hypothesis that in a generally covariant quantum theory the physical time-flow is not a universal property of the mechanical theory, but rather it is determined by the thermodynamical state of the system (*thermal time hypothesis*)".

If thus one considers that the time involved in the second and third phases of the cosmic evolution as represented in figure 2, namely, from point β to point ω is such *a thermal time* denoted by $\tau$ , all the energy densities except $\rho_\Lambda$ that, as shown in [1], can be considered as an integration *constant*, depend on the thermal time: they can be called *comoving* energy densities. According to this interpretation, rather than $\rho_\Lambda$, it is $\rho_c$ itself that can be called the time-dependent dark energy density because, being by its very definition – see Eq. (8) –the surface energy density of the Hubble horizon, it is a dark negative pressure and goes asymptotically to $\rho_\Lambda$. With such a terminology, one can rewrite the time dependent flatness sum rule as

$$\rho_{\mathrm{M}}(\tau)+\rho_{\mathrm{DM}}(\tau)+\rho_\Lambda=\rho_c(\tau)=\rho_{\mathrm{DE}}(\tau)=\Lambda_{\mathrm{eff}}(\tau)/8\pi G_N \quad \text{with } \Lambda_{\mathrm{eff}}(\infty)=\Lambda \tag{12}$$

where the index M stands for the sum of the contributions to the energy density of pressure-less matter and radiation.

### *4.4 The entanglement thermodynamics*

In the important article named *Spacetime thermodynamics without hidden degrees of freedom* [13], the authors stress the difference between ordinary statistical thermodynamics and what they call *entanglement thermodynamics*. Quantum entanglement induces entropy even in a pure state of a pair of entangled degrees of freedom living on both sides of a horizon when one integrates out one of the two degrees of freedom. Whereas in statistical thermodynamics, entropy measures our ignorance, in entanglement thermodynamics, entropy measures quantum entanglement, namely a “genuinely quantum phenomenon which does not exist in classical physics”. They also stress that the entropy used by T. Jacobson [14] to reinterpret the Einstein’s equation as an equation of state is also an entanglement entropy rather than a statistical entropy.

### *4.5 Holography and the thermodynamics of the Hubble horizon*

#### *4.5.1 The Bekenstein' constant, the cosmological constant and the thermodynamics of the Hubble horizon*

Coming back to the proposed interpretation of the dark matter as an anti-de Sitter world matter, we see that the flatness sum rule gets a new *holographic* meaning: at all epochs, namely from point β to point ψ, the cosmology of the universe that emerges from the quantum vacuum is modeled in terms of a (thermal) time-dependent energy density of the *quantum holographic Hubble world-matter* that depends on three of the foundational constants of [1], the quantum of entropy $k_{\mathrm{B}}$ and the quantum of spacetime area $A_{\mathrm{P}}$ whose ratio is the Bekenstein constant $B=\dfrac{k_B}{4A_P}$, and the cosmological constant $\Lambda$. It is holographic in the sense that it equates the total energy density in the bulk, which is a *volume* energy density, with the total negative pressure exerted on the boundary of the bulk, which is a *surface* energy density. Eq.(12) equates the sum total of the energy densities in the volume (including the integration constant $\rho_{\Lambda}$) to the surface energy density of the Hubble horizon equal to the negative isotropic pressure.

The expression of the Hubble world-matter energy density exhibits a genuine complementarity relation between an anti-de Sitter density corresponding to the past event horizon (point α in figure 2), and a de Sitter density corresponding to the future event horizon (point ω in figure 2), through the analytic continuation introduced above.

When the cosmological term $\rho_{\Lambda}$ and the matter energy density $\rho_{\mathrm{M}}(\tau)$ are put in the right hand side, defining $-\rho_{\mathrm{AdS}}(-\Lambda_{\mathrm{eff}}(\tau))$ as $\rho_{\mathrm{DM}}(\tau)$ and $\rho_{\mathrm{dS}}(\Lambda_{\mathrm{eff}}(\tau)$ as $\rho_{\mathrm{DE}}(\tau)=\rho_{c}(\tau)-\rho_{\Lambda}-\rho_{\mathrm{M}}(\tau)$ eq. (12) becomes the light cone kinematical identity $E = pc$ (where $p$ is the four momentum leading to a pressure $P$) that reads

$$\rho_{\mathrm{DM}}(\tau)=\rho_{\mathrm{DE}}(\tau)\ \text{or} -\rho_{\mathrm{AdS}}(-\Lambda_{\mathrm{eff}}(\tau))\equiv\rho_{\mathrm{dS}}(\Lambda_{\mathrm{eff}}(\tau)), \tag{13}$$

which means $P_{\text{AdS}} + P_{\text{dS}} = \rho_{\text{AdS}} + \rho_{\text{dS}} = 0$ .

Such a complementarity relation cannot be but a consequence of a quantum process. Actually, about the transition that occurs at point β, there is an assumption which is often implicit in the literature, and that we want to make explicit here: namely, that during the primordial inflation stage, has occurred the breaking of the fermion-antifermion symmetry and that the following expansion of the universe is driven by the fermions that have survived to the global annihilation of fermions and antifermions.

#### *4.5.2 Qualitative aspect of the large scale' structure of the Hubble world-matter*

Holography allows understanding qualitatively the structure of the distribution at large scale of galaxy clusters which can be expected in the ΛCDM model and which is observable in computer simulations: a web of tension-less $(P_{\text{AdS}} + P_{\text{dS}} = 0)$ filaments of world-matter, with a length of the order of the Hubble radius *L*, with a mass density of order $\rho_{\text{dS}} \approx L / L^3$ (holography), freely moving (comoving) in the de Sitter spacetime associated with the cosmological constant $(\rho_\Lambda + P_\Lambda = 0)$

To understand more quantitatively such a structure, one can make use of the third complementarity schema, the one of the gauge-gravity duality.

### *4.6 Gauge-gravity duality in QCD*

#### *4.6.1 The trace anomaly and the QCD world-matter*

If one considers the content of the present-day universe in the perspective of its emergence from the quantum vacuum, one is automatically led to focus on QCD since baryonic matter is made of QCD quanta, namely *u* and *d* quarks and gluons. About dark matter, one thus wants to know what does become the flatness sum rule at point δ in figure 2, namely when emerges the colorless universe out of the QCD vacuum at the confinement/deconfinement transition.

At this transition it is valuable to interpret Eq. (12) in a way that makes explicit the contribution of the QCD quantum vacuum: since the matter and dark matter are supposed to emerge from this vacuum, one can associate their energy densities in the QCD world-matter. To complete the content of the universe, one must add to the contributions of the QCD quanta the ones of the non-QCD or electroweak quanta which will lead, in the present-day budget, to the negligible radiation contribution and to the cosmological term $\rho_\Lambda$ . These non-QCD quanta result from the weak or electromagnetic decays of the very numerous hadrons (mainly pions) produced at the confinement-deconfinement transition at point δ.

Eq. (12) thus becomes:

$$\rho_{\mathrm{QCD}}^{\mathrm{WM}} + \rho_{\mathrm{EW}}^{\mathrm{WM}} = \rho_{\mathrm{c}} \tag{14}$$

Where just as $\rho_\Lambda$ in the present day budget, $\rho_{\mathrm{EW}}^{\mathrm{WM}}$ plays the role of an integration constant at point $\delta$ .

As an unbroken non-abelian renormalizable gauge theory, QCD is asymptotically free at short distance, and singular at large distance. The QCD Lagrangian, without quarks or with massless quarks i.e. in the so-called chiral limit, is scale invariant since the coupling constant is dimensionless, but quantization leads to a non-vanishing vacuum expectation of the trace of the stress energy tensor (the so-called trace or conformal anomaly), a spontaneous symmetry breaking. In the cosmological context, one can assume that the quantum of the scalar field $\phi$ of Eq. (1) could play the role of the Nambu-Goldstone boson associated with this spontaneous symmetry breaking.

The dynamical breaking of scale invariance, called dimensional transmutation after Coleman and Weinberg [15] is apparent in the fact that "the renormalization has replaced a one-parameter family of unrenormalized theories, characterized by their values of the dimensionless unrenormalized gauge coupling $g_0$, by a one-parameter family of renormalized theories, characterized by their value of the dimension-one *scale mass* $\mathcal{M}(g,\mu)$ "[16]. This scale mass $\mathcal{M}(g,\mu)$ is said to be renormalization group invariant: it satisfies the Callan [17]-Symanzik [18] differential equation

$$\left[\mu\frac{\partial}{\partial\mu} + \beta(g)\frac{\partial}{\partial g}\right]\mathcal{M}(g,\mu) = 0 \tag{15}$$

With

$$\beta(g) = -\frac{1}{2} b_0 g^3 + O(g^5) \tag{16}$$

Where the constant $b_0 = 11N_c - 2N_f$ ($N_c$ is the number of colors, and $N_f$ the number of quark flavors) appears in the lowest order (one loop) perturbative calculations of radiative corrections, but also appears in the non-perturbative regime of QCD. In effect, the constant $b_0$ figures as the coefficient of the essential singularity at zero coupling constant in the expression of the scale mass $\mathcal{M}(g,\mu) = \mu \exp\left(-1/b_0 g^2\right)$.

This scale mass enters the definition of QCD vacuum condensates such as a (fermionic) quark-antiquark condensate $\left(\langle \overline{q}q \rangle_0 \propto \mathcal{M}(g,\mu)^3\right)$ or a (bosonic) gluonic condensate $\left(\left\langle F^{a}_{\mu\nu} F^{a\mu\nu} \right\rangle_0 \propto \mathcal{M}(g,\mu)^4\right)$, and the constant $b_0$ appears as a multiplicative factor in the contribution of all the vacuum condensates in the trace anomaly. For instance, the contribution of the gluon condensate to the trace anomaly reads:

$$\left\langle T^{\mu}_{\mu} \right\rangle_0 = -\frac{1}{8}\left[11N_c - 2N_f\right]\left\langle \frac{\alpha_s}{\pi}\left(F^{a}_{\mu\nu} F^{a\mu\nu}\right)^r \right\rangle_0 \tag{17}$$

Where the symbol $\langle \ldots \rangle_0$ stands for vacuum expectations in flat space-time integration, and $\alpha_S = g^2/4\pi$.

### *4.6.2 The QCD vacuum condensates and the gauge theory-superconductor analogy*

In [16], S. Adler has developed a gauge-theory superconductor analogy that can be used to equate the trace anomaly with the world matter density. In the case of QCD, the fermionic and bosonic condensates which are proportional to powers of the scale mass are the analog of the electron pair in a superconductor proportional to the energy gap, and the slowly varying metric (represented in our interpretation by the dilaton field $\phi$) that perturbs the condensates in the vacuum is the analog of the slowly varying electromagnetic field that perturbs the electron pair condensate in the superconductor.

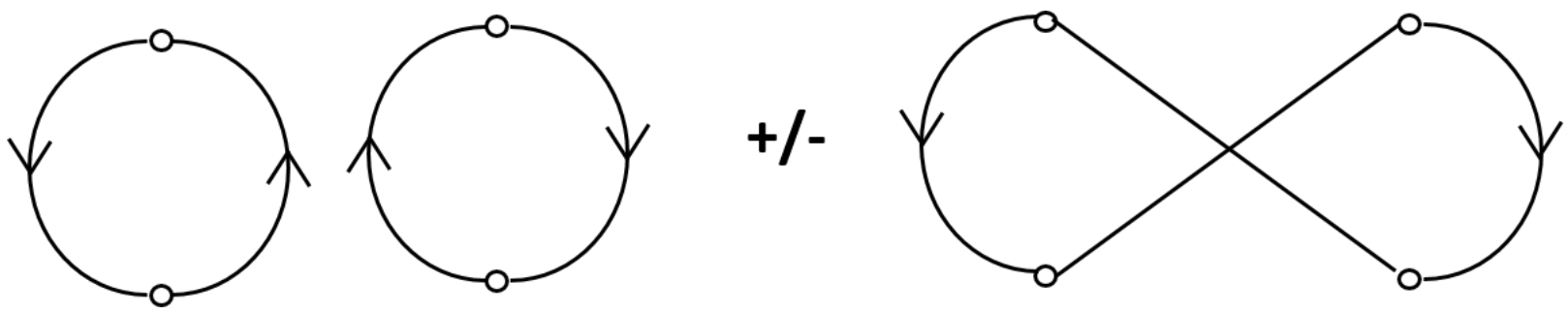

Figure 3a. Vacuum-vacuum quantum statistics: (Bose-Einstein + sign) or (Fermi-Dirac- sign) correlations

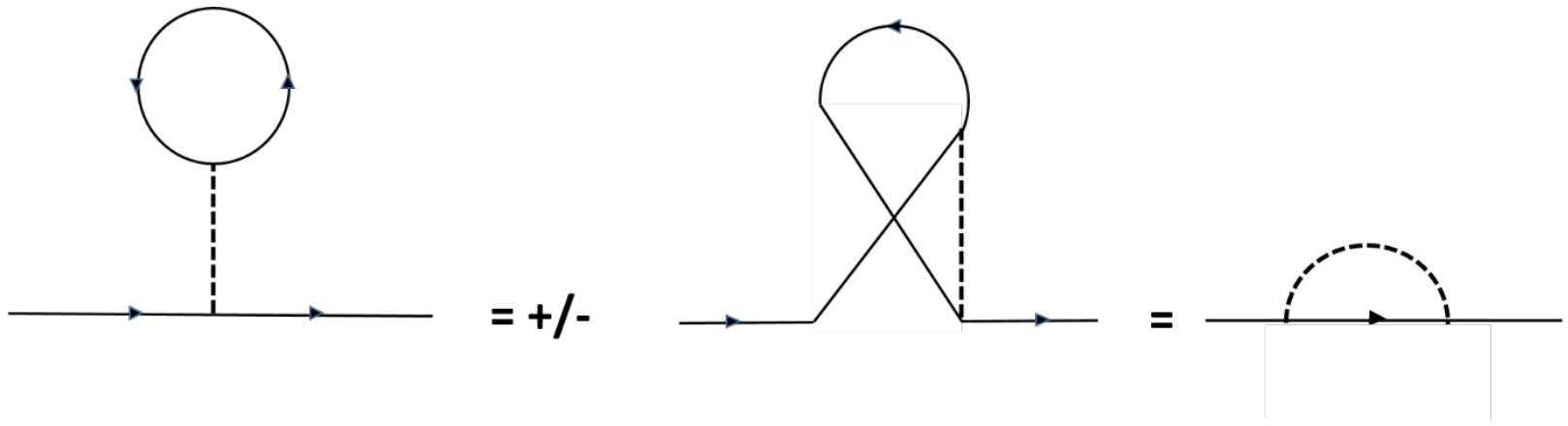

Figure 3b. A "tadpole" diagram in which a boson (resp. fermion) exchanges a virtual dilaton with a vacuum loop involving a particle identical to it, is transformed trough the interchange of identical particles, into a positive, i.e. increasing the mass (resp. negative, i.e. increasing the energy) self-energy diagram.

The minus sign in the right hand side shows that when the constant $b_0 = 11N_c - 2N_f$ is positive, all the QCD condensates contribute negatively to the energy density, which means that the QCD world-matter is globally an anti-de Sitter world-matter (dominance of an anti de Sitter world-matter over a smaller de Sitter world-matter) . This multiplicative factor $b_0$ allows reading, thanks to the well-known property that boson and fermion loops contribute in quantum field theory with opposite signs (see figures 3a and 3b), the relative contributions of the components of the QCD vacuum to the full world-matter:

- the bosonic (gluon) loops, proportional to $N_c$, contribute to the anti-de Sitter world matter which represents the potential negative gravitational energy density of the gluons and will become the dark matter, and
- the fermionic (quark) loops, proportional to $N_f$, contribute to the normal de Sitter world matter, which, per our interpretation, represents the kinetic energy density of the quarks which have survived to the global annihilation of fermions and antifermions, namely the constituents of the baryonic matter

We thus see that with $N_c$, the number of colors equal to 3, and the number $N_f$ of light flavors of quarks (u, *d, s*) that form the baryonic matter, also equal to 3, the ratio of the bosonic to fermionic loops in the QCD quantum vacuum contributing to the world-matter is 5.5, which is in excellent agreement with the ratio of the dark matter to the baryonic matter energy densities in ΛCDM. The superconductor analogy was used in [19] by Nielsen and Olesen who proposed a suggestive model in which the analog of the QCD vacuum is a superconductor of type II involving unconfined chromo-magnetic monopoles moving freely along magnetic flux lines that form a "three-dimensional pattern which resembles spaghetti".

More recently, the superconductor analogy is used to model dark matter by means of a Bose-Einstein Condensation (BEC) mechanism. For instance, in [20] in which the *axion*, the Nambu-Goldstone scalar associated with the Peccei-Quinn solution to the strong CP problem [21] is assumed to be the dark matter particle that condenses in the potential induced in QCD at the color confinement scale and acquires a time (or temperature) dependent mass. In this model, the properties (mass and coupling to 2 γ) must be fine-tuned to explain why such a particle has not been discovered yet.

Unlike this axion-like model, the model that I propose is a gluon BEC model that does not need any ad hoc scalar field: the dilaton, the quantum of the scalar field $\phi$ related to the determinant of the metric, acts as the quasi-particle with a time (or temperature) dependent mass representing, the collective gravitational effect of the Bose-Einstein gluon condensate.

## 5 Two short remarks to conclude

My first remark takes the form of two questions about the Lochak' monopoles.

1. Can the baryons be considered as Lochak's massive magnetic monopoles in the framework of QCD at large number of colors *N*? Such an interpretation had been proposed by Ed. Witten [22]

> "Indeed, the baryon mass is of order *N*, which can be written as 1/(1/*N*). But 1/*N* is the "coupling constant" of the strong interactions, which characterizes the interaction among mesons. 1/*N* plays in QCD roughly the role that $\alpha$ plays in spontaneously broken gauge theories of the weak and electromagnetic interactions. The fact that the baryon mass is of order 1/(1/*N*) is analogous to the fact that the Polyakov-'t Hooft monopole mass is of order $1/\alpha$."

2. Could not the pattern of Nielsen Olesen [19] chromo-magnetic flux lines in which the chromo-magnetic monopoles (Lochak-Witten monopoles?), i.e. baryons move freely resemble the one of dark matter filaments appearing in the simulations of the distribution of galaxies at large scale?

My second remark is an invitation to read or re-read the book edited as a tribute to Louis de Broglie for his sixtieth birthday [23] which contains extremely interesting contributions, of which I extract this quotation of Schrödinger

> One must regard the "observation of an electron" as an *event* that occurs within a train of de Broglie waves when a contraption is interposed in it which by its very nature cannot but answer by discrete responses: a photographic emulsion, a luminescent screen, a Geiger counter. And one must, to repeat this, hold on the wave aspect throughout. This includes, that the equations between frequencies and frequency differences, the resonance condition that governs throughout, must not be multiplied by Planck's constant h and then interpreted as tiny energy balances of microscopic processes between tiny specks of something that have, to say the least, no permanent

> existence (…)These inconsistencies will be avoided by returning to a wave theory that is not continually abrogated by dice-miracles; not of course of the naïve wave theory of yore, but to a more sophisticated one, based on second quantization and the non-individuality of "particles"[24]

**Acknowledgement.** It is a pleasure to thank the de Broglie Institute, and particularly Georges Lochak, for the stimulating discussions we had and the interest they manifested for my work at the meeting on the 4th of June 2019.

---